\documentclass[pdflatex,sn-mathphys-num]{sn-jnl}

\usepackage{graphicx}%
\usepackage{multirow}%
\usepackage{amsmath,amssymb,amsfonts}%
\usepackage{amsthm}%
\usepackage{mathrsfs}%
\usepackage[title]{appendix}%
\usepackage{xcolor}%
\usepackage{textcomp}%
\usepackage{manyfoot}%
\usepackage{booktabs}%
\usepackage{algorithm}%
\usepackage{algorithmicx}%
\usepackage{algpseudocode}%
\usepackage{listings}%

\usepackage[percent]{overpic}
\usepackage{siunitx}

\theoremstyle{thmstyleone}%
\theoremstyle{thmstyletwo}%

\theoremstyle{thmstylethree}%

\begin{document}

\title{Rapid axial loading of a grating MOT with a cold-atom beam}

\author[1]{\fnm{Rachel} \sur{Cannon}}
\author[1]{\fnm{Aidan S} \sur{Arnold}}
\author[1]{\fnm{Paul F} \sur{Griffin}}
\author[1]{\fnm{Erling} \sur{Riis}}
\author*[1]{\fnm{Oliver S} \sur{Burrow}}\email{oliver.burrow@strath.ac.uk}

\affil[1]{\orgdiv{Department of Physics}, \orgname{University of Strathclyde}, \orgaddress{\street{107 Rottenrow East}, \city{Glasgow}, \postcode{G4 0NG}, \country{UK}}}


\abstract{Laser-cooled atoms are increasingly being used to realise practical quantum devices, motivating the development of compact and robust atom sources. Grating magneto-optical traps (gMOTs) simplify the cold-atom source architecture but are typically vapour-loaded and provide limited atomic flux. Here we explore the loading of gMOTs from cold-atom beams. We numerically simulate loading to show that unbalanced diffracted beams deflect incoming atoms away from the trap centre, thereby strongly constraining radial loading. In contrast, axial loading injects atoms directly into the trapping volume and largely avoids these effects. We experimentally demonstrate rapid axial loading of a gMOT, achieving loading rates of $2.1 \times 10^9$ atoms~s$^{-1}$ using a moving optical molasses to transfer atoms from a 2D MOT into the gMOT. These results establish axial loading as a robust route to high-flux gMOT operation for portable cold-atom systems.}


\maketitle

\section{Introduction}\label{sec1}

Laser-cooled atoms provide well-controlled atomic platforms for precise and stable measurements. 
These capabilities underpin a growing range of practical cold-atom quantum technologies, from sensing applications such as atomic clocks \cite{Takamoto2020_clock,pelle2022cold,Ascarrunz2018_SpectraDynamics,Bregazzi2024_clock} and atom interferometers \cite{bongs2019taking,bidel2020absolute,Seo2026_atomInt} to emerging quantum computing \cite{henriet2020quantum,wintersperger2023neutral} and memory systems \cite{mol2023quantum}. 
The intrinsic stability of atom-based measurements makes them attractive for position, navigation and timing (PNT) applications, where classical sensors are limited by long-term drift \cite{bongs2019taking}.
In this context, quantum sensors are being developed to complement global navigation satellite systems (GNSS), on which a significant proportion of modern economic activity is reliant upon \cite{Blackett2018}.
To realise these technologies in field applications, a key sub-system, the cold-atom source must become portable, requiring components with low size, weight and power (SWaP) budgets.

Grating magneto-optical traps \cite{nshii2013surface} (gMOTs) simplify the laser cooling of atomic ensembles to an architecture requiring a single cooling beam and a planar diffractive optic, providing large optical access and making them well suited to reducing the SWaP of cold-atom-based sensors.  
These gMOT devices have enabled laser cooling of rubidium atoms for demonstrations of compact atomic clocks \cite{elvin2019cold,lewis2022grating,Bregazzi2024_clock}, atom interferometers \cite{Seo2026_atomInt,lee2022compact,Alex2026_HARLEQUIN}, optical lattice structures \cite{bregazzi2025single}, Bose-Einstein condensates \cite{calviac2025bose,heine2024high}, and ultracold electron sources \cite{franssen2019compact}. Furthermore, grating MOTs have successfully laser cooled a variety of other atomic species in lithium \cite{barker2019single}, strontium \cite{sitaram2020confinement} and caesium atoms \cite{xu2025dual,Bregazzi2026_Cs_gMOT}. 

Grating MOT devices are typically loaded from atomic vapour, providing atom loading rates on the order of 10$^{8}$~s$^{-1}$ with equilibrium atom numbers of several 10$^{7}$ \cite{McGilligan15_phaseSpace}. In these vapour-loaded systems, the cold-atom loading dynamics are dictated by the Rb partial pressure. Higher Rb pressures enable faster atomic loading but increase the background collision rate, which shortens the coherence time of any prepared quantum state and limits achievable measurement contrast.

The traditional solution to this trade-off is straightforward: atoms are pre-cooled in a high-pressure Rb environment and then transferred into a low-pressure region, decoupling the loading rate from the background pressure. This approach also benefits the steady-state atom number, since collisional losses, dominant in vapour-loaded systems, are largely eliminated in a separate cold-atom source. Commonly used methods for improving the atomic flux include 2D and 2D$^{+}$ MOTs, low-velocity intense sources (LVIS), and Zeeman slowers \cite{riis1990atom,Dieckmann1998twodmot,Lu1996_LVIS,Phillips1982_Zeeman}.

Loading a gMOT with a cooled atomic beam has been demonstrated using both a Zeeman slower approach \cite{barker2019single,sitaram2020confinement} and 2D MOT configurations \cite{calviac2025bose,heine2024high,imhof2017two}. In recent work, Ref. \cite{heine2024high} achieved loading rates up to 10$^9$~atoms~s$^{-1}$ using a 2D\textsuperscript{+} configuration, where atoms enter the grating MOT region on an axis perpendicular to the grating normal (radial loading).
This improved the atomic loading rate by an order of magnitude upon previous work \cite{imhof2017two}, where a 2D gMOT loaded a gMOT.  A key factor for this improvement is control of the trajectory and speed of the atoms entering the gMOT \cite{heine2023high}.  As gMOT optics are made of binary gratings, in addition to the light diffracted inwards to form the MOT beams, they also diffract light outwards, and these beams interact with and deflect incoming cold atom beams.  

In this work, we experimentally demonstrate axial loading of a grating magneto-optical trap (gMOT), where atoms are loaded along the grating normal, using a 2D+ MOT inspired loading scheme. We also present a simple model showing how radiation-pressure-induced deflections of atomic trajectories influence loading from atomic beams and limit radial loading.  Ultimately this means that gMOTs have a minimum capture velocity in addition to the traditional maximum capture velocity of a regular six-beam MOT.  The experimental consequence is that radially loaded gMOTs have favoured trajectories and speeds best achieved using a 2D$^+$ MOT.  

The approach we have taken here is to bypass these deflections and load the gMOT axially through a hole in the gMOT optic.  We show why axial loading is less sensitive to experimental parameters than radial loading via simulation, describe an experimental apparatus that implements this axial loading method, and then present loading rate results  of $2.1 \times 10^9$~atoms~s$^{-1}$. The axial transfer is mediated by a moving optical molasses formed between the push beam and the incident gMOT beam.   Ultimately, utilising the gMOT beam as part of the 2D$^+$ MOT beam geometry simplifies the apparatus by removing the technical challenge of installing a mirror within the vacuum chamber.

\section{Methods}
\subsection{Average force simulation of atomic trajectories in gMOTs}

With an intensity-balanced, well aligned traditional six-beam MOT, there are no unbalanced beams that deflect atoms away from the trapping volume.  A beam of atoms can enter the trapping volume without a significant hindrance.

\begin{figure}[ht]
    \centering

    \begin{minipage}[t]{0.49\linewidth}
        \centering
        \begin{overpic}[width=\linewidth]{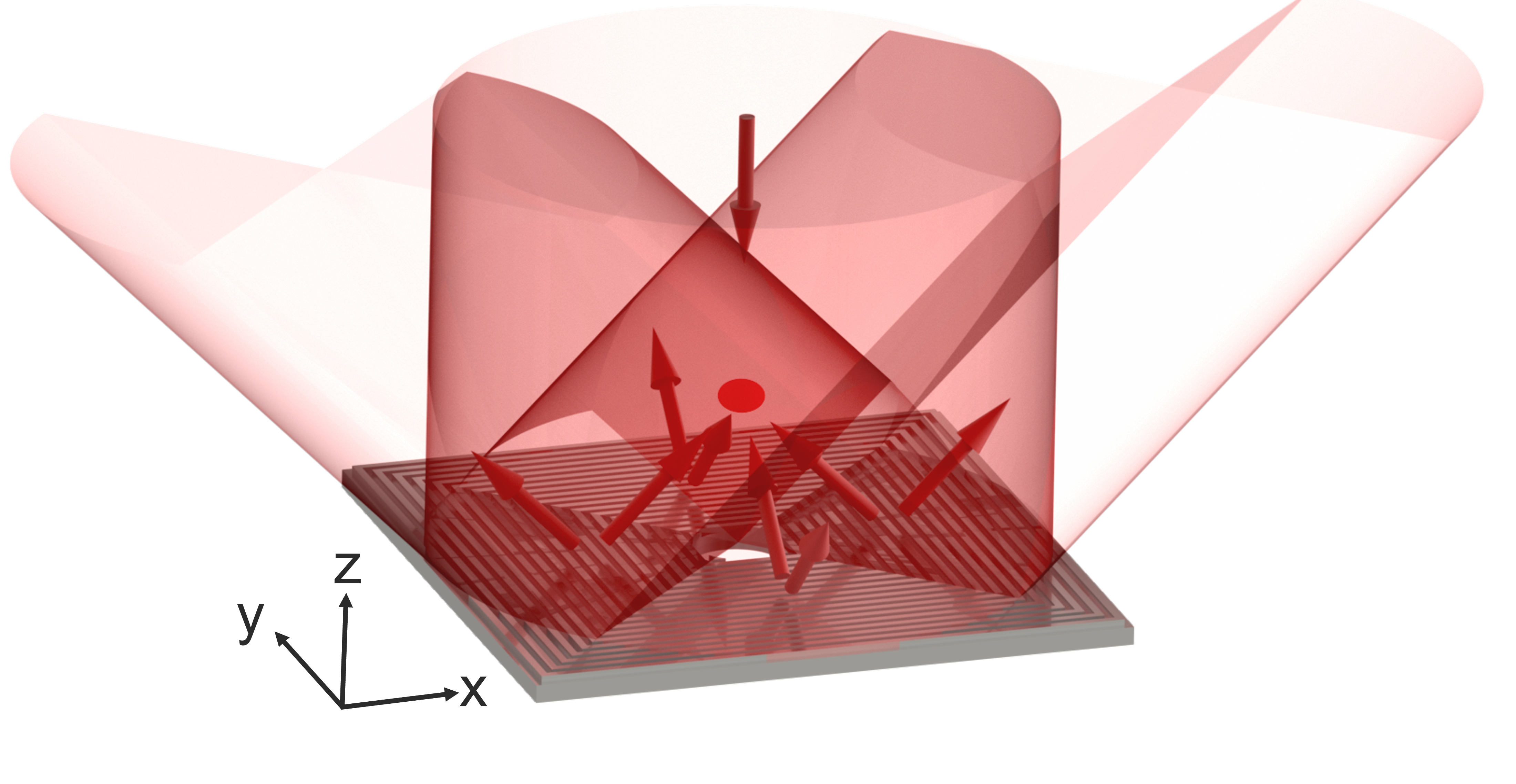}
            \put(3,57){\textbf{(a)}}
        \end{overpic}
    \end{minipage}\hfill
    \begin{minipage}[t]{0.49\linewidth}
        \centering
        \begin{overpic}[width=\linewidth]{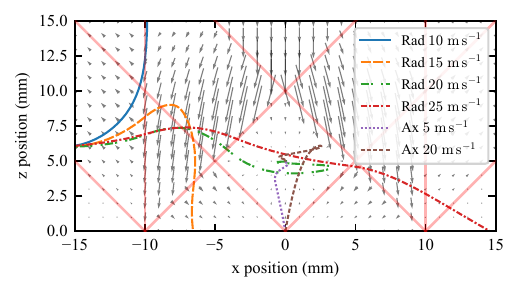}
            \put(3,57){\textbf{(b)}}
        \end{overpic}
    \end{minipage}

    \caption{(a) Illustration of a QUAD gMOT grating, with the input and diffracted beams in the $x-z$ plane illustrated. Arrows indicate wavevectors of light input and diffracted on the grating in three dimensions, with an ellipsoid representing the position the MOT forms. The gMOT optic has a 3 mm diameter hole in its centre. The symmetry of the QUAD-style chip lends itself for a simplified simulation here. (b) Simulated trajectories for atoms launched radially (Rad) at 10–25~$\mathrm{m\,s^{-1}}$ overlaid on the gMOT average force vectors, evaluated at the arrow midpoint. Slow atoms are deflected by the diffracted beams, while faster atoms overfly the trap. 5 and 20~$\mathrm{m\,s^{-1}}$ axial (Ax) trajectories with a $\mp10^\circ$ angle illustrate how axial loading avoids these transverse deflections. Solid red lines demarcate the regions between different beam overlap geometries.}
    
    \label{fig:Fig1_QuadandTraj}
\end{figure}

In contrast, outside of a gMOT’s trapping volume, the beams are not balanced; atoms are acted upon by incomplete beam geometry, and are often pushed away from the trap region rather than into it.  
In addition to the diffracted beams contributing to the optical geometry, the gMOT optic produces outward-diffracted beams that do not contribute to trapping; instead these beams push atoms away from the trap region, even beyond the confines of the input beam.
Consequently, loading a gMOT from a cold-atom beam requires atoms to lie within a trajectory-dependent velocity band: fast enough to overcome deflections from incomplete beam geometries, yet slow enough to remain within the capture velocity of the gMOT \cite{heine2023high}.
These deflections can be completely avoided by loading atoms axially through the grating. A central aperture in the grating provides a direct path into the trapping region. Such apertures have previously been shown not to degrade gMOT performance \cite{bregazzi2021simple}.

To illustrate these effects, we performed trajectory simulations using a heuristic force model for a two-level toy atom ($F=0 \rightarrow F'=1$) \cite{Vangeleyn2009_tetra}, otherwise corresponding to the parameters of $^{87}$Rb. The simulations were restricted to the $x$–$z$ plane, taking advantage of the symmetry of the QUAD-style gMOT chip (Fig.~\ref{fig:Fig1_QuadandTraj}(a)), where the relevant beams for this 2D simulation are illustrated. The QUAD and TRI gMOT geometries are defined in Ref.~\cite{burrow2023optimal}. The choice of QUAD rather than TRI grating geometry, and the symmetry axis used for radial loading, are discussed at the end of this section. Stochastic scattering forces were neglected. At each timestep the simulation determined which laser beams illuminated the atom, and therefore the forces acting on it. The force calculation included contributions from the incident and zeroth-order beams, the $\pm 1$ diffracted orders, and within the trapping region, an additional positive-$z$ term representing the effect of the $y$-axis beams.

The scattering force from each beam was calculated using the standard rate-equation expression:

\begin{equation}
\mathbf{F}_i = 
\hbar \mathbf{k}_i\,\frac{\Gamma}{2}\,\frac{I_i}{I_\text{sat}}
\left[
\begin{aligned}
&\frac{I_\pi}{
    1+\tfrac{I_\text{tot}}{I_\text{sat}}
    +4\left(\tfrac{\Delta - \mathbf{k}_i\cdot \mathbf{v}}{\Gamma}\right)^2} \\
&+ \frac{I_{\sigma^-}}{
    1+\tfrac{I_\text{tot}}{I_\text{sat}}
    +4\left(\tfrac{\Delta - \mathbf{k}_i\cdot \mathbf{v} + \delta_B}{\Gamma}\right)^2} \\
&+ \frac{I_{\sigma^+}}{
    1+\tfrac{I_\text{tot}}{I_\text{sat}}
    +4\left(\tfrac{\Delta - \mathbf{k}_i\cdot \mathbf{v} - \delta_B}{\Gamma}\right)^2}
\end{aligned}
\right].
\end{equation}

Here, $\Gamma$ is the natural linewidth, $\Delta$ the laser detuning, $I_i$ the intensity of beam $i$, and $I_\text{tot}$ the summed intensity from all beams. The Doppler shift is included via the projection of the atomic velocity $\mathbf{v}$ onto the wavevector $\mathbf{k}_i$. The Zeeman shift is given by $\delta_B = g_F \mu_B B/\hbar$, with $g_F = 1$ for the ground state used here. 

The fractional $\pi$ and $\sigma^{\pm}$ components of each beam were determined from the angle $\theta$ between $\mathbf{k}_i$ and the local magnetic field, as defined below:
\[
I_\pi = \tfrac{1}{2}\sin^2\theta, \quad  
I_{\sigma^-} = \tfrac{1}{4}(1-\cos\theta)^2, \quad  
I_{\sigma^+} = \tfrac{1}{4}(1+\cos\theta)^2.
\]

The magnetic field was calculated using Magpylib's implementation of the Biot-Savart law \cite{Ortner2020magpylib} for a pair of anti-Helmholtz coils with 15 mm radius and 25 mm separation, with the coil current chosen to generate the desired quadrupole gradient along $z$, $B_z' = \partial B_z / \partial z$, at the trap centre. The total scattering force was then obtained as $\mathbf{F} = \sum_i \mathbf{F}_i$ and used to propagate atomic trajectories forward in time with a Runge-Kutta method (RK45) implemented in the SciPy library \cite{2020SciPy-NMeth}.

To illustrate the effect of minimum and maximum capture velocities, we simulated atoms launched in the $x$-direction, 6~mm above the grating, with initial velocities of 10, 15, 20, and 25~$\mathrm{m\,s^{-1}}$. Figure~\ref{fig:Fig1_QuadandTraj} (b) shows their trajectories overlaid on a force vector map, where the arrows represent the local radiative force an atom at rest would experience. The simulation used a diffraction efficiency of 25\% into the first order and 0\% into the zeroth order, with a diffraction angle of $45^{\circ}$. The incident intensity was set to a uniform $I/I_{\mathrm{sat}}=5$, the magnetic field gradient was $B_z' = 20$~G/cm with the centre 5~mm above the grating, and the detuning was fixed at $-1.66~\Gamma$. Each trajectory begins at the point where the atom first enters the diffracted beam at this height.

At 10~$\mathrm{m\,s^{-1}}$ atoms are deflected before reaching the incident beam,  whilst at 15~$\mathrm{m\,s^{-1}}$ the incident beam pushes them into the grating.  At 20~$\mathrm{m\,s^{-1}}$ atoms are successfully captured, while at 25~$\mathrm{m\,s^{-1}}$ they pass through the trap without being captured.  These example trajectories highlight the existence of both a minimum and a maximum capture velocity: atoms that are too slow are deflected away before reaching the trap, while those that are too fast pass through without being confined. Meanwhile, axially loaded atoms are also illustrated at both 5 and 20~$\mathrm{m\,s^{-1}}$, demonstrating the capture of atoms over a wide velocity range.

\begin{figure}
    \centering
    \includegraphics[width=0.5\columnwidth]{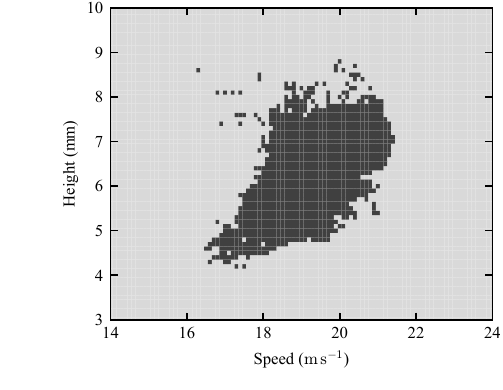}
    \caption{Binary greyscale representation of trapping success in a radially loaded gMOT versus loading speed and input height (dark grey: trapped; light grey: not trapped).    }
    \label{LoadingSuccess_height}
\end{figure}

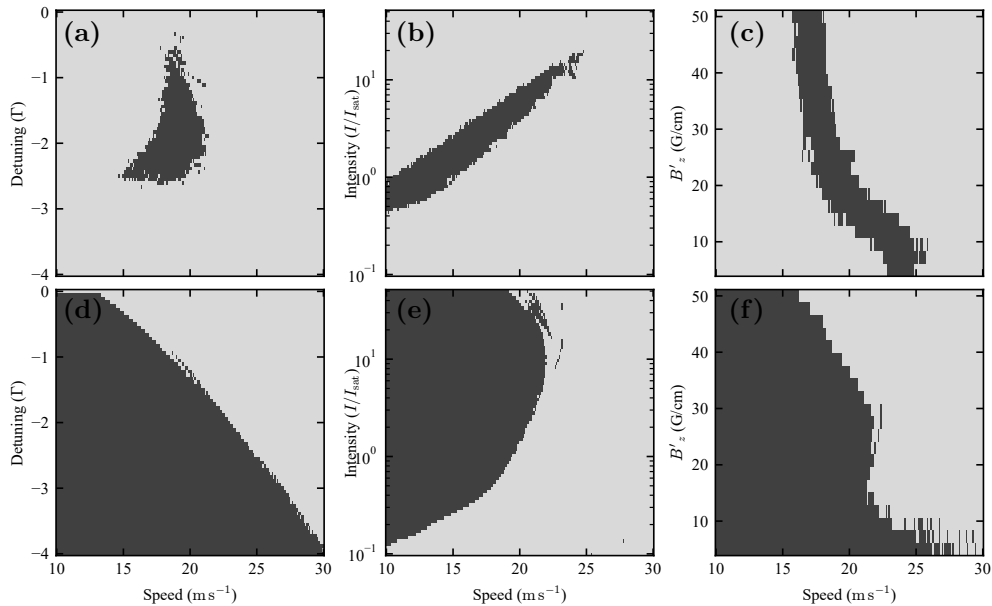
\begin{figure}[ht]
\centering
\begin{overpic}[width=\linewidth]{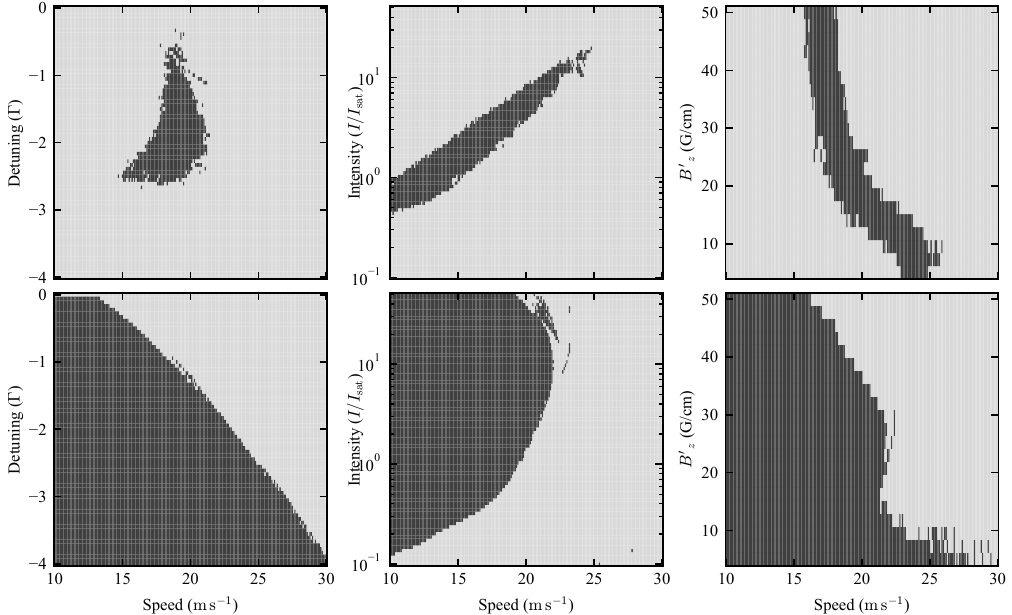}
  \put(6,57){\textbf{(a)}}
  \put(6,29){\textbf{(d)}}

  \put(39.5,57){\textbf{(b)}}
  \put(39.5,29){\textbf{(e)}}

  \put(73,57){\textbf{(c)}}
  \put(73,29){\textbf{(f)}}
\end{overpic}

\caption{Binary greyscale maps of simulated trapping success in a gMOT under radial (top row) and axial (bottom row) loading. Columns left to right show trapping as a function of loading speed versus laser detuning (a,d), intensity (b,e), and axial magnetic field gradient (c,f), respectively. Dark grey regions indicate successful trapping, while light grey regions indicate no trapping.}
\label{LoadingSuccess_six}
\end{figure}

Although such individual trajectories provide an intuitive picture of the capture dynamics, they do not provide a complete view of how the capture probability depends on atomic velocity and experimental parameters. To address this, we performed a systematic set of simulations in which atoms were launched into the gMOT simulation across a range of input velocities and ranges of experimental parameters.  At the end of 100~$\mathrm{ms}$ of simulated time evolution, atoms located within the trap centre were labelled as trapped, while all others were labelled as not trapped. These results are presented in Figs. \ref{LoadingSuccess_height} and \ref{LoadingSuccess_six}. We examined three key experimental parameters in turn: the incident intensity, the detuning and the axial magnetic field gradient $B_z'$. For each case, simulations were performed for atoms loaded both radially and axially; in addition, for radial loading we also varied the input height above the grating.

The simulations reveal a clear difference between the two loading geometries. Loading axially consistently yields a much larger parameter space of successful captured atoms compared to radial loading, which is restricted to a narrow velocity band. Consequentially axial loading is considerably less sensitive to experimental parameters when compared to radial-loading. For this radial-loading, Fig.~\ref{LoadingSuccess_height} shows that the input height must be controlled to within $\pm 1.5$~mm, and the acceptable velocity width is only about 5~$\mathrm{m\,s^{-1}}$.
The dependence on intensity and detuning reinforces this picture. As shown in Fig.~\ref{LoadingSuccess_six}~(a),(b),(d) and (e), increasing intensity or detuning broadens the capture range for axial loading, whilst radial loading has a narrow acceptable velocity range.
The behaviour with magnetic field gradient is somewhat different. As seen in Fig.~\ref{LoadingSuccess_six}~(c) and (f), axial loading maintains a wide capture range across large variations in $B_z'$, with slightly increased tolerance at lower gradients.  At high gradients, atoms that pass the trap centre can be Zeeman shifted out of resonance with the cooling light, reducing the slowing efficiency. Again the radially loaded trap remains constrained to a narrow velocity window.

These results show that radial loading requires tight control over both the input velocity and the trajectory of the atom beam. Dieckmann et al \cite{Dieckmann1998twodmot} have demonstrated that a 2D$^+$ MOT can achieve this, producing atomic beams with velocity distributions of order 4~$\mathrm{m\,s^{-1}}$ FWHM, with control over the central frequency, comparable to the narrow acceptance width identified here. By contrast, axial loading of the gMOT is far more forgiving, with successful capture observed over a broad velocity range and with little sensitivity to alignment. This motivates a closer examination of axial loading configurations, including the use of a push beam to control the flux into the trap.  

As this work targets future compact vacuum encapsulation, a QUAD grating was chosen because its Pd coating tolerates higher bake temperatures than the Al coating typically used on TRI gratings \cite{burrow2021stand}. Whilst TRI gratings have a larger capture volume, this is less relevant here because the MOT is loaded from a cold-atom beam.

The QUAD grating provides two radial symmetry extremes for loading: either directly opposing an outwardly diffracted beam or along a seam of the grating. 
In this work we simulate loading against a diffracted beam. In earlier experiments with a 2D gMOT source, we observed that radial loading along the symmetry aligned with the slowing beam produced a factor of 2–3 higher loading rate than loading along the seam.
We speculate that loading against the outwardly diffracted beam enables the capture of faster atoms, which was better suited to the broad velocity distribution produced by the 2D MOT source used in those experiments. For a narrower velocity distribution cold-atom source, such as that produced by a 2D+ MOT, both radial symmetries may perform comparably, albeit with different preferred mean atom velocities.

\subsection{Compact Challenges}
To achieve a compact system, two design considerations were taken into account. The first concerned the proximity of the 2D and 3D MOT magnetic fields. The 3D MOT requires a point-centred quadrupole magnetic field to enable cooling in all three dimensions, whereas the 2D MOT desires a radial quadrupole only. 
The two magnetic fields will add vectorially, and bringing them closer together increases the distortion of each from its desired form. In principle, this superposition need not be detrimental if both MOTs share a common symmetry axis, as the combined field then preserves the intended on-axis structure. However, any deviation from this shared-axis symmetry, such as tilting or offsetting the 2D MOT axis relative to the gMOT laser-beam axis, introduces unwanted cross-talk between the fields.

\begin{figure}[h!]
    \centering

    \begin{minipage}{0.54\columnwidth}
        \centering
        \begin{overpic}[width=\linewidth]{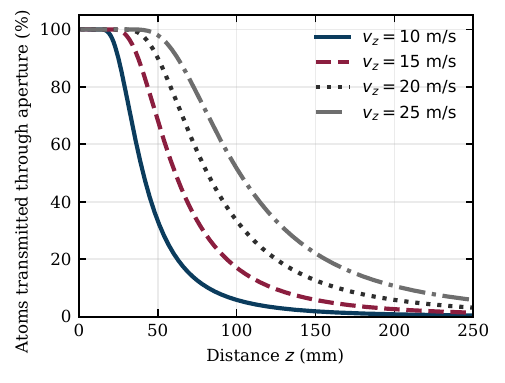}
            \put(3,81){\textbf{(a)}}
        \end{overpic}
    \end{minipage}
    \hfill
    \begin{minipage}{0.44\columnwidth}
        \centering
        \begin{overpic}[width=\linewidth]{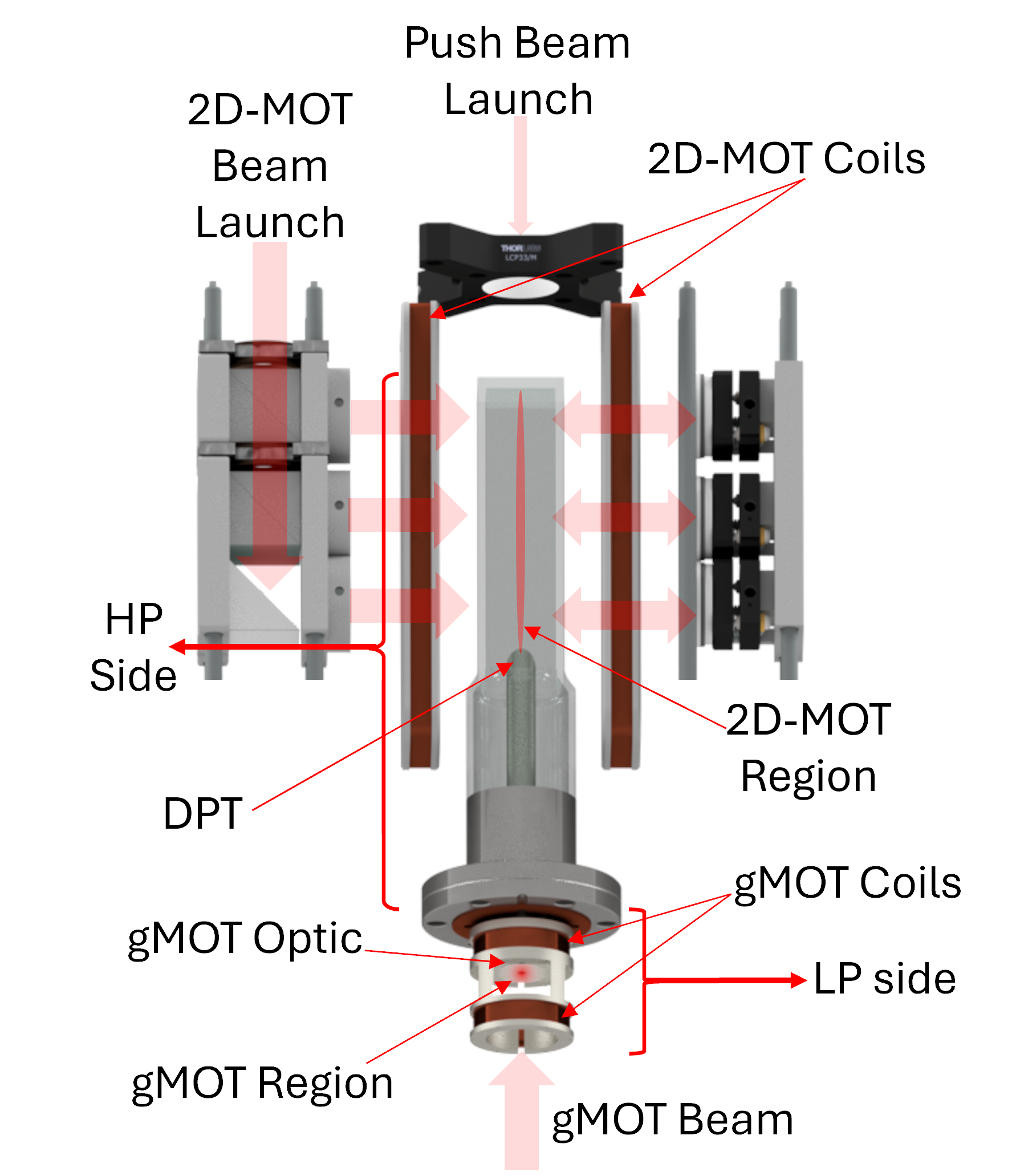}
            \put(3,95){\textbf{(b)}}
        \end{overpic}
    \end{minipage}

    \caption{(a) The fraction of atoms that will transmit through a 3~mm diameter aperture at a distance $z$, for different axial beam velocities of 10, 15, 20 and 25~$\mathrm{m\,s^{-1}}$ is calculated by modelling the beam's radial velocity as a Maxwell-Boltzmann distribution. (b) A schematic illustration of the 2D MOT loaded gMOT apparatus, annotated to show how the 2D MOT beams are delivered, the location of the 2D and 3D MOT coils, and the differentiation between the high and low pressure (HP and LP) sides of the vacuum. The differential pumping tube (DPT), gMOT coil former and diffractive optic mount is a single piece of 3D printed titanium, to facilitate the transfer of atoms between the HP and LP sides of the vacuum.}
    
    \label{fig:Fig3_transAndApparatus}
\end{figure}

The second consideration was what fraction of a diverging cold-atom beam would pass through the aperture in the grating at a given distance.
Increasing the distance between the 2D MOT and the aperture would reduce atomic flux.

One method to mitigate this issue would be to increase the aperture in the gMOT optic.  However, increasing the aperture diameter would compromise the gMOT optic's capability; previous studies of gMOT optics with apertures saw a reduced atom number for aperture diameters above 3~mm \cite{bregazzi2021simple}.  
By modelling the radial velocity distribution with Maxwell–Boltzmann statistics, we calculated the fraction of atoms that pass through the fixed aperture for different axial velocities.  Typically 2D MOTs produce atomic beams with divergences of ~30~mrad, commensurate with a radial temperature near the Doppler limit \cite{Dieckmann1998twodmot,Schoser2002intensesource,Chaudhuri2006realization,Castagna2006Novel}, and our model indicated gMOT capture velocities between 10 and $20~\mathrm{m\,s^{-1}}$. The following values were used as parameters in the statistical model: the radial velocity corresponds to \SI{150}{\micro\kelvin} for $^{87}$Rb, and the longitudinal velocity of 10, 15, and   $20~\mathrm{m\,s^{-1}}$, for an aperture of 3~mm diameter.    Illustrated in Fig. \ref{fig:Fig3_transAndApparatus}(a), we observe how the cold-atom flux is reduced with increased distance between the 2D MOT and the aperture.

Informed by this statistical model, magnetic systems were designed for both MOTs using the Python package Magpylib \cite{Ortner2020magpylib}. For the gMOT, a pair of anti-Helmholtz-like circular coils generate the target quadrupole field centred on the MOT location and $z$-axis.  For the 2D MOT a radial quadrupole along the z-axis was produced using two orthogonal pairs of rectangular coils, each in an anti-Helmholtz-like configuration.  These coils are illustrated in Fig. \ref{fig:Fig3_transAndApparatus} (b) This 2D MOT current geometry cancels the axial field along the central z-axis.

There is a trade-off in the separation between the MOTs. Reducing the distance increases the atomic flux, as a larger fraction of atoms pass through the apertures between the chambers. However, bringing the magnetic systems closer together increases the influence of the gMOT quadrupole field in the 2D MOT region, particularly if the field axes are not perfectly aligned. To reduce this cross-talk, the gMOT coils were designed for compact in-vacuum operation so that their stray field decays rapidly with distance. The coils were therefore enclosed within a standard DN40CF cube to minimise their size.

The apparatus described in the following section reflects the interplay of this design trade-off and the decision to use off-the-shelf components during this exploratory stage of the work.

\subsection{Apparatus}

The vacuum design consists of a two-chambered vessel with a high-pressure (HP) 2D-MOT side and a low-pressure (LP) 3D gMOT side, illustrated in Fig. \ref{fig:Fig3_transAndApparatus} (b). The 2D MOT side is comprised of a glass cell of outer dimensions 100~mm x 30~mm x 30~mm, with 2.5~mm thick glass, where a cooled atomic beam is created through 2D cooling of $^{87}$Rb atoms. The atoms are transferred through a differential pumping tube to the gMOT side, housed in a DN40CF cube. 
The glass cell used provides the necessary optical access and vacuum performance for a 2D MOT, and was identified as the most compact commercially available solution. However, the cell's glass–metal join imposes a fixed length that limits how closely the two MOT regions can be positioned. 
A single 3D-printed titanium piece was designed to integrate the coil former for the 3D-MOT in-vacuum coils, a platform for the gMOT optic and the differential pumping tube in a unified structure, as shown in Fig \ref{fig:Fig3_transAndApparatus} (b).  The 3D coils were wound using kapton coated wire, and performed optimally at a magnetic field gradient of 38~G/cm. The incorporated differential pumping tube was 105~mm long, with a 3~mm aperture along its centre.  
The titanium piece was mounted on a machined solid-gasket between the LP and HP volumes. SAES rubidium dispensers were placed on the HP side, while only the LP side was actively pumped to support future miniaturisation of the design. The differential pumping ratio is 20:1, with a calculated conductance of 0.5~l/s.

The 3D beam launch optics expand, circularly polarise, and collimate the output of a polarisation maintaining optical fibre, delivering an 80~mW beam of $1/$e$^2$ radius of 15~mm to the gMOT optic. 
The 'QUAD' gMOT optic is 20 mm $\times$ 20~mm, and was purchased from Kelvin Nanotechnology. The HP side has two windows for optical access, where a camera and photodiode are placed orthogonal to the grating axis at each window for measurements of the MOT loading and atom number. 

A partial illustration of the combined 2D–3D MOT assembly is shown in Fig. \ref{fig:Fig3_transAndApparatus} (b), with only one transverse axis of the 2D MOT rendered for clarity.  The 2D MOT coils are two pairs of anti-Helmholtz rectangular coils, which performed optimally when each pair produced 18~G/cm.
The 2D MOT optics are housed in custom-designed 3D-printed PLA mounts, aligned in a custom cage-mount configuration. The 2D beam launch delivers a $1/$e$^2$ beam radius of 10~mm and approximately $60~\mathrm{mW}$ in optical power per axis. Each beam is directed to a linear chain of two polarising beam splitters (PBS) and a right-angled prism for reflection at each, to produce three beams in each axis, as seen in Fig. \ref{fig:Fig3_transAndApparatus} (b) \cite{ramirez2006multistage}. Half-waveplates placed before each PBS control the power directed at the atoms, and are adjusted for even power distribution. Circular polarisation of each beam is achieved using quarter-waveplates mounted in custom designed 3D prints. Each beam is subsequently retro-reflected by a mirror to form the 2D MOT. The push beam is launched to be axially aligned to the centre of the glass cell. The push beam has a $1/$e$^2$ beam radius of 2.4~mm and was found to have no preferred state of linear polarisation. 

\section{Results}

The system performance is characterised by the atom number in the gMOT, $N$, and the corresponding initial loading rate, $\left.\frac{dN}{dt}\right|_{t=0}$. The atom number is measured via fluorescence detected on an avalanche photodiode. The photodiode current is converted to atom number using a calibration based on the detector quantum efficiency, collection solid angle (assuming isotropic emission), and the calculated atomic scattering rate for the applied laser intensity and detuning.

\begin{figure}[t]
\centering

\begin{minipage}[t]{0.49\columnwidth}
    \centering
    \vspace{0pt}
    \includegraphics[width=\linewidth]{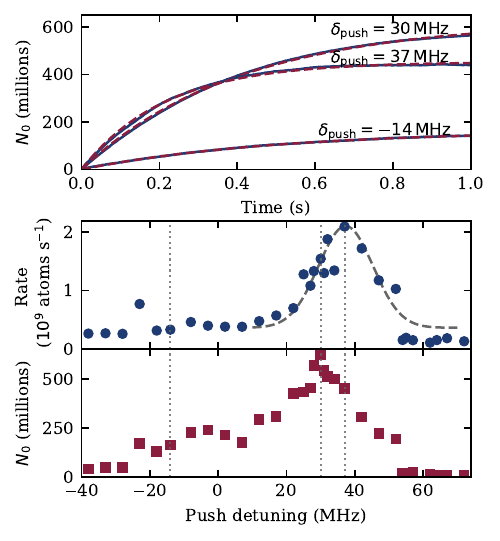}
    
    \setlength{\unitlength}{1mm}
    \begin{picture}(0,0)
        \put(-20,68){\textbf{(a)}}
        \put(-20,42){\textbf{(b)}}
        \put(-20,25){\textbf{(c)}}
    \end{picture}
\end{minipage}
\hfill
\begin{minipage}[t]{0.49\columnwidth}
    \centering
    \vspace{0pt}
    \includegraphics[width=\linewidth]{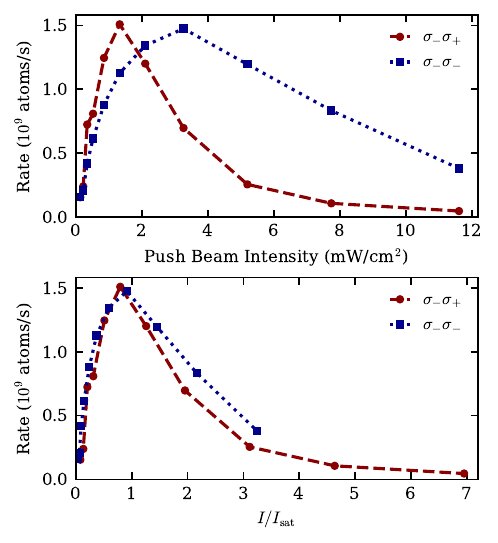}
    
    \vspace{1.5em} 
    \setlength{\unitlength}{1mm}
    \begin{picture}(0,0)
        \put(-33,76){\textbf{(d)}}
        \put(-33,43){\textbf{(e)}}
        \end{picture}
\end{minipage}

\caption{(a) Average loading curves (blue) for selected push-beam detunings, each obtained from an average of ten loading measurements. The dark red dashed lines show exponential fits of the form $N(t) = N_0 \left(1 - e^{-t/\tau}\right)$. The detunings shown correspond to the vertical dotted lines in (b) and (c). (b) Loading rate $R = N_0/\tau$ as a function of push-beam detuning. The maximum rate of $2.1 \times 10^9$ atoms\,s$^{-1}$ occurs at $37$\,MHz. The dashed curve is a guide to the eye. (c) Equilibrium atom number $N_0$ extracted from the same fits. The maximum atom number of $6.2 \times 10^8$ Rb atoms occurs at $30$\,MHz. (d) 2D-MOT loading rate as a function of push-beam intensity for the two circular polarisation configurations, $\sigma_{-}\sigma_{+}$ and $\sigma_{-}\sigma_{-}$. In(e), the polarisation configurations intensities are normalised to their respective saturation intensities \cite{SteckRb87}.}

\label{Fig_data_combined}
\end{figure}
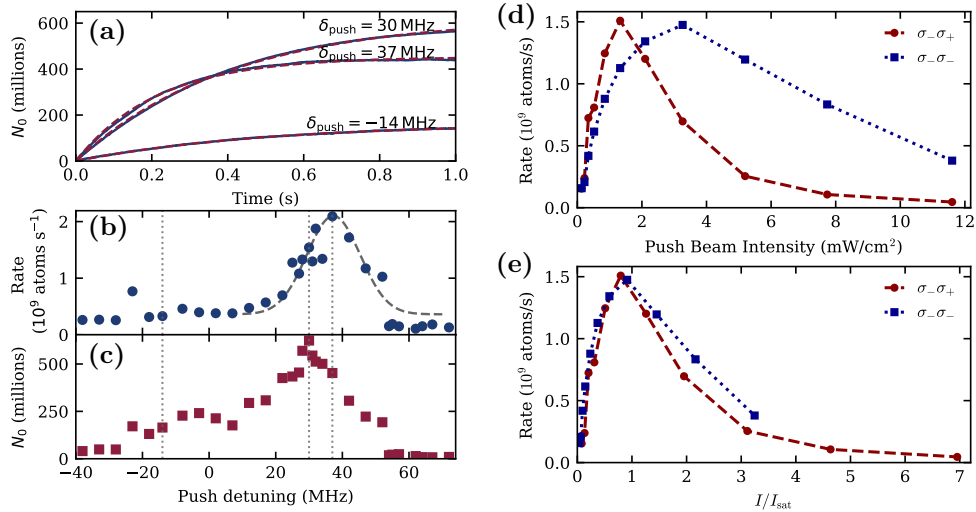

The experimental performance is extracted from fits of the form 
$N(t) = N_0 \left(1 - e^{-t/\tau}\right)$.
The resulting loading curves are shown in Fig.~\ref{Fig_data_combined}~(a), with representative values summarised in Table~\ref{tab:gmot_performance} for three push-beam detunings corresponding to equal detuning to the gMOT, maximum atom number, and maximum loading rate. The loading dynamics are strongly dependent on the push-beam detuning.

\begin{table}[h]
\centering
\small
\setlength{\tabcolsep}{4pt}
\caption{Summary of fit parameters extracted from Fig.~\ref{Fig_data_combined} (a), showing the equilibrium atom number, initial loading rate, and corresponding moving molasses velocity for representative push-beam detunings.}
\begin{tabular}{lccc}
\hline
 & gMOT Freq & Max $N_{\mathrm{eq}}$ & Max loading rate \\
\hline
Push Detuning (MHz) & $-14$ & $30$ & $37$ \\
$N_{\mathrm{eq}}$ & $1.3 \times 10^{8}$ & $6.2 \times 10^{8}$ & $4.5 \times 10^{8}$ \\
Loading Rate (s$^{-1}$) & $3.2 \times 10^{8}$ & $1.5 \times 10^{9}$ & $2.1 \times 10^{9}$ \\
Molasses velocity ($\mathrm{m\,s^{-1}}$) & $0$ & $17.2$ & $19.9$ \\
\hline
\end{tabular}
\label{tab:gmot_performance}
\end{table}

These results show that the push beam-detuning which maximises loading rate does not coincide with that which maximises equilibrium atom number, with differences between the two optima at the $\sim 30\%$ level.
Operation with the push beam set equal to the gMOT detuning yields substantially reduced performance, demonstrating that independent control of the push-beam frequency is essential.

This behaviour can be understood in terms of a moving-molasses interaction between the counter-propagating push and incident gMOT beams. 
A frequency difference between the two beams produces a velocity-dependent imbalance in the scattering forces \cite{riis1990atom}, such that atoms are carried along in the moving frame where the two frequencies are equal. The velocity of this moving frame is given by $v = \frac{\lambda \, \Delta f}{2}$, with $\lambda$ the wavelength of the light and $\Delta{f}$ the frequency difference between the counter-propagating beams. 
To laser cool, the gMOT beam must be negatively detuned from the $F=2 \rightarrow F'=3$ transition of the $^{87}$Rb D$_2$ line. We therefore investigated how the push-beam detuning, referenced to the same transition, affects the loading rate. The 3D beam detuning and 2D beam were fixed to -14~MHz and -10~MHz, respectively, from the cooling transition. The power of the 3D beam was 100~mW, giving an intensity of 14.2~mW/cm$^2$ in the 3D MOT, LP side. The push beam intensity was kept consistent at 5.2~mW/cm$^2$, giving a saturation parameter of 1.45 for an $I_{\mathrm{sat}}$ of 3.57~mW/cm$^2$.

Figure \ref{Fig_data_combined}~(b) shows the dependence of the gMOT loading rate on the push-beam detuning, and Fig \ref{Fig_data_combined}~(c) the equilibrium atom number dependence on push beam detuning. As the detuning is increased to the blue of the cooling transition, the loading rate shows a well-defined peak centred at a push detuning of approximately 37~MHz. Interpreting this behaviour within the framework of a moving-molasses interaction, this optimal detuning corresponds to an atomic-beam velocity of 19.9~$\mathrm{m\,s^{-1}}$, assuming equal intensities for the counter-propagating beams. 
A Gaussian curve has been drawn on Fig \ref{Fig_data_combined} (b), which has a full-width half maxima of around 23 MHz, indicating the working velocity range is 9~$\mathrm{m\,s^{-1}}$.

While the observed detuning dependence is consistent with a moving-molasses picture, this interpretation raises questions regarding the underlying force dynamics. In a simple average-force model, the radiation pressure experienced by an atom depends dispersively on its velocity relative to the moving frame defined by the two beams, and would therefore be expected to produce a strong sensitivity of the gMOT loading rate to velocity. The comparatively weak dependence observed here suggests that additional mechanisms act to mitigate this velocity dispersion, overcoming the heating naively expected from such a dispersive molasses. 

To further probe the mechanism underlying atom transfer, the polarisation of the push beam was modified by inserting a quarter-waveplate into the beam path. This allowed the effects of circular polarisation to be explored, motivated by the possibility that the counter-propagating push beam and gMOT light could form a 1-D optical lattice. The circular polarisation quoted here is using the conventional definition of the rotation direction of the electric field relative to the propagation direction.\footnote{Note that this is in contrast to the convention widely adopted in the laser cooling community early on where the circular polarisation was defined relative to a fixed quantisation axis \cite{Dalibard:89} and hence making it more obvious to determine the transitions induced between magnetic sublevels. Under that convention, the configuration denoted here as $\sigma^{-}/\sigma^{-}$ is the field configuration of a 1D MOT colloquially referred to in the laser cooling community as corkscrew polarisation.}
For the following data the polarisation of the incident gMOT beam remained fixed and is defined as $\sigma^{-}$ with respect to the propagation direction.

 Fig.~\ref{Fig_data_combined}~(d) shows the measured atomic loading rate as a function of push-beam optical power for two circular-polarisation configurations: $\sigma^{-}$$\sigma^{+}$ and $\sigma^{-}$$\sigma^{-}$. In both cases, comparable loading behaviour is observed. The primary difference between the datasets is a scaling of the required optical power, which is consistent with the differing saturation intensities associated with isotropic and stretched-state transitions. Using our chosen definition of polarisation, the $\sigma^{-}$$\sigma^{+}$ configuration describes a standing wave which is considered circular at every point in space, optically pumping the atoms into the stretched state at a lower effective saturation intensity. The opposite polarisation configuration is thought of as linear everywhere, with an atom travelling the wave therefore requiring a higher saturation intensity than with the standing wave.  When scaling the intensity by the associated saturation intensity, as can be seen in Fig.~\ref{Fig_data_combined}~(e), peak loading rate occurs at the same scaled intensity value, indicating that changing the push-beam polarisation primarily modifies the fluorescence rate rather than the underlying beam-formation mechanism.

The persistence of the loading behaviour under changes in relative polarisation suggests that the observed atom transfer cannot be attributed solely to a simple optical-lattice mechanism, which would be expected to be highly sensitive to the polarisation configuration of the counter-propagating fields. Instead, these observations point toward a more complex cooling process. In particular, polarisation-gradient cooling mechanisms are known to trap atoms at temperatures of order a hundred micro-Kelvin, for which the associated restoring forces can readily compete with or exceed the average radiation-pressure force \cite{Dalibard:85}. It is therefore not unexpected that such a mechanism could remain effective in the presence of the velocity-dependent forces discussed above. A detailed treatment of this regime is beyond the scope of the present work, and we conclude by attributing the atom-transfer process to a polarisation-gradient, moving-molasses-type interaction.  Unlike a pure optical-lattice mechanism, which would be highly sensitive to the relative polarisation of the counter-propagating beams, the observed transfer remains operative under changes in polarisation, indicating operational robustness to polarisation drifts in practical implementations.

\section{Discussion}

The results presented here demonstrate a substantial enhancement in both the equilibrium atom number and atomic flux delivered to the gMOT, compared to vapour-loaded traps \cite{McGilligan15_phaseSpace,burrow2021stand}. Both this work and Ref. \cite{heine2024high} achieve an order-of-magnitude improvement over earlier 2D-MOT loaded gMOT experiments using rubidium atoms \cite{imhof2017two}. These enhancements are attributed to improved control over the velocity distribution of atoms entering the gMOT region, enabled by the addition of a `plus'-type beam. In our work, this high performance is achieved by axially loading the gMOT from a novel 2D$^+$-type MOT, where the incident gMOT beam provides the counter-propagating plus beam. 

While achieving comparable atom numbers are obtained to those reported for radially loaded gMOT configurations \cite{heine2024high}, the axial approach demonstrated here provides a complementary route that is particularly well suited to compact and portable implementations.  A key practical advantage of axial loading is its reduced sensitivity to experimental alignment. In previous radial implementations of atomic beam loaded gMOTs, the loading rate was found to depend strongly on the orientation of the outwardly diffracted beams relative to the incident cold-atom beam. The average-force simulations provide a physical explanation for this behaviour and indicate that axial loading is intrinsically less sensitive to the trajectory of the atomic beam than radial loading schemes. 
In addition, the intrinsic symmetry of the axially loaded gMOT magnetic fields minimises magnetic-field cross-talk as the system is scaled down. These simulations also provide a framework for understanding geometric acceptance in the gMOT loading process. While the present model classifies trajectories as either trapped or not trapped, it could be extended via Monte Carlo sampling over an assumed velocity distribution to generate relative flux predictions and enable direct comparison with experiment. Further extensions incorporating internal atomic states and stochastic effects would improve predictive power but are beyond the scope of the present work.

For quantum sensing applications operating near the quantum projection noise limit, the increased flux demonstrated here will increase in the total number of atoms detected per unit time $N_{total}$, and hence will improve their sensitivity scaling by $\sqrt{N_{total}}$. These improvements are achieved without requiring elevated Rb vapour pressure in the gMOT vacuum region, thereby minimising collisional decoherence of the atomic state during the measurement phase.

Beyond sensing applications, the enhanced atom number also results in a significant increase in the achievable optical density of the trapped ensemble. The ability to generate high optical densities in a compact geometry therefore opens a pathway towards portable cold-atom quantum memories and light–matter interfaces for quantum communication\cite{saglamyurek2018autlertownes}.  Additionally, the combination of high atom number, low pressure, and large optical access may also be advantageous for fundamental quantum measurement experiments.  For example, in devices such as quantum gas microscopes, atoms are typically transported to a separate science chamber to achieve sufficient optical access, at the cost of atom number and cycle time, motivating efforts to realise architectures that reduce or eliminate this transport \cite{Buob2024_GasMicroscope}.  In platforms such as neutral atom quantum computing, the techniques demonstrated here could pair well with approaches such as meta-optics, which are being explored to generate large arrays of optical traps within compact, reduced-SWaP architectures \cite{Fang2025_arrays}.

There remains scope for future performance improvements of the apparatus. For example, Ref. \cite{calviac2025bose} demonstrates a factor of ten improvement in atom number when implementing a flat-top beam profile over a Gaussian profile for the incident gMOT beam. Applying a similar beam-shaping approach in the presented system could therefore provide a straightforward route to increase further the loading of the gMOT. Overall, these results demonstrate a simple and robust method for significantly enhancing the performance of a gMOT-based cold-atom source. By combining high atom number, increased flux, and intrinsic compatibility with compact geometries, this approach provides a strong foundation for the development of next-generation portable cold-atom systems for sensing and quantum technologies.

\backmatter

\bmhead{Acknowledgements}

The authors thank Hendrik Heine for valuable discussions on the loading dynamics of grating magneto-optical traps, which motivated the approach taken in this work.  The authors also thank Ernst M. Rasel for valuable discussions.

\section*{Declarations}

\bmhead{Funding} 
This work was supported by Innovate UK under the SBRI programme through CPI-TMD [Grant number 10072054].

\bmhead{Conflicts of interest} 
The authors declare that they have no competing interests.

\bmhead{Data availability} 

The datasets supporting the findings of this study will be deposited in the University of Strathclyde PURE repository. A persistent DOI will be assigned and included upon publication.

\bmhead{Author contributions} 
R.C. performed the experimental work, analysed the data, co-designed the experiment and co-wrote the manuscript.
O.S.B. conceived and co-designed the experiment, secured funding, performed the simulations, and co-wrote the manuscript. A.S.A., P.F.G., and E.R. contributed to the analysis and interpretation of the physics, with guidance on the simulations from A.S.A. and on the experimental work from P.F.G. and E.R. All authors reviewed and refined the manuscript.

\bibliography{RachelBib.bib}

\end{document}